\documentclass[11pt,article]{iopart}
\usepackage{microtype}
\usepackage{array}
\usepackage{caption}
\usepackage{braket}
\usepackage{bbm}
\usepackage[english]{babel}
\usepackage{color}
\usepackage{soul}
\usepackage{dcolumn}
\usepackage{bm}
\usepackage{color}
\usepackage{overpic}
\usepackage{latexsym}
\usepackage{epstopdf}
\usepackage{appendix}
\newcommand{\mathd}{\mathrm{d}}
\def\bf{\textbf}
\newcommand{\text}[1]{\mbox{\footnotesize{#1}}}
\newcommand{\eqref}[1]{(\ref{#1})}
\newcommand{\ketbra}[2]{\ket{#1}\!\bra{#2}}

\begin{document}
\title[]{\textbf{Improving the precision of frequency estimation via long-time coherences}}

\author{A Smirne$^1$, A Lemmer$^1$, M B Plenio$^1$, S F Huelga$^1$}
\address{$^1$ Institute of Theoretical Physics and IQST, Universit{\"a}t Ulm, Albert-Einstein-Allee 11, 89069 Ulm, Germany}

\ead{andrea.smirne@uni-ulm.de}

\begin{abstract}
In the last years several estimation strategies have been formulated to determine
the value of an unknown parameter in the most precise way, taking into account the presence of noise.
These strategies typically rely on the use of quantum entanglement
between the sensing probes and they have been shown
to be optimal in the asymptotic limit in the number of probes,
as long as one performs measurements on shorter and shorter time scales.
Here, we present a different approach to frequency estimation,
which exploits quantum coherence in the state
of each sensing particle in the long time limit and is obtained by
properly engineering the environment. By means of a commonly used master equation, we show that our strategy can overcome the precision achievable with entanglement-based strategies for a finite number of probes.  We discuss a possible implementation of the scheme in a realistic setup that uses trapped ions as quantum sensors.
\end{abstract}
%
%
%
%
%
%
%
%
%
%
%
%
%
%

\section{Introduction}
How precisely can we estimate the value of an unknown parameter?
The answer to this question provides us with a paradigmatic example of how
quantum features
can lead to a significant advantage over any classical strategy.
In classical experiments with $N$ sensing particles, i.e. $N$ probes,
the central limit theorem sets the mean-squared-error scaling
of the best estimation strategies to $N^{-1}$, according to the standard quantum limit (SQL).
Instead, the ultimate quantum limit, or Heisenberg limit (HL),
achieves a $N^{-2}$ scaling of the error,
which can be in principle reached via the preparation of entangled probes \cite{Giovannetti2004,Toth2014,Demkowicz2015}.

The advantage due to quantum estimation strategies
is jeopardized by the interaction of the probing systems
with the surrounding environment, potentially reducing the improvement
to a constant factor \cite{huelga1997,Escher2011,Demkowicz2012,Demkowicz2015}.
To overcome this limitation, in recent years several approaches have been put forward,
relying on non-negligible spatial \cite{Dorner2012,Jeske2014} or temporal
\cite{Jones2009,Matsuzaki2011,chin2012,Macieszczak2015,Smirne2016,Latune2016,Jarzyna2017,Yousefjani2017,Haase2018}
correlations in the environment, as well on a
particular geometry of the system-environment coupling \cite{Chaves2013},
possibly allowing for error correction techniques \cite{chiara2000,preskill2000,Sushkov2014,Duer2014,Arrad2014,Unden2016,Sekatski2017,Demkowicz2017,Zhou2018} or fault tolerant strategies \cite{datta2018}.
These approaches are mostly focused on achieving the best
asymptotic scaling of the estimation precision with respect to the number of probes,
especially aiming at surpassing the SQL, thus also providing a clear fingerprint of the quantum origin
of the obtained enhancement.
For example, the preparation of initially entangled states can
increase the estimation precision in the presence of time-correlated noise
as long as the
system is interrogated at times short enough to minimize the impact of decoherence.
Measurements at shorter and shorter time intervals with increasing $N$
guarantee the optimal estimation strategy in the asymptotic regime $N \rightarrow \infty$,
achieving a scaling of the error as $N^{-\alpha}$ with values of $\alpha$ strictly greater than 1 \cite{Jan2018}.

On the other hand, the preparation of a high number of entangled probes
and the access to short interrogation times is certainly too demanding in several situations
of interest \cite{adesso2016,friis2017,adesso2018}. For this reason, it is important to obtain a deeper understanding of estimation strategies which take into account
the finite resources at hand in a more realistic way and, in particular,
the finite (and possibly small) number of probes,
as well as a minimum duration time of each experimental run \cite{munro2018,munro2017}.
In this paper, we show that
one can design a simple and effective frequency estimation strategy, which
does not call for the preparation of entangled probes and relies on measurements in the long-time
regime of the probes' evolution. Most importantly, for finite values of $N$ this strategy can lead to an enhanced precision with respect to the
entanglement-based one; indeed, the latter will still be the optimal strategy
in the asymptotic limit of $N$ \cite{chin2012,Macieszczak2015,Smirne2016}.

Our approach exploits a coherent effect in the dynamics of the probes, which is known in the literature as coherence trapping (CT).
As suggested by the name, CT is manifest by the presence of a non-negligible amount of coherence
in the stationary state of a system, despite its interaction with the surrounding environment.
CT is characteristic of some specific spectral densities, such as the super-ohmic spectral form within a pure dephasing model
\cite{luczka1990,reina2002,morozov2012,addis2014},
but it can also be induced by engineering a part of the environmental degrees of freedom
\cite{man2015}, or due to initial system-environment correlations \cite{zhang2015}.
Quite remarkably, recently the presence of nonzero asymptotic quantum coherence
has been shown for the spin-boson model
in the presence of a generic (neither parallel nor transversal) coupling \cite{Guarnieri2018}.
Note that the emergence of CT is different from the generation of a so-called decoherence free subspace \cite{Lidar2003,Vanier2005},
since the coherences of the system need not to be invariant under the dynamics,
but simply to converge to a finite value.
CT has been already studied extensively in the literature, but, to our knowledge,
this is the first investigation of its possible use for enhanced metrology.
Here, we focus on the dynamics of a qubit coupled to a damped harmonic oscillator, which provides us with the simplest conceptual framework
for the demonstration of our proposal
and,
indeed, describes many physical systems of interest \cite{haroche2006}, which can be realized with different technologies
in typically controllable setups, such as atoms confined in optical cavities or trapped ion arquitectures.

After illustrating how CT can emerge by engineering the environment of the model at hand and how this can be used to improve
estimation precision for finite values of $N$, we will show that these ideas can be implemented in current setups that employ trapped ions as quantum sensors.

\noindent

\section{Standard metrological bounds}
\subsection{General framework for frequency estimation}
We consider the task of estimating a frequency $\omega$, within the so-called frequentist (or Fisher-Information) approach \cite{Demkowicz2015}.
$N$ probes are prepared in a possibly entangled state $\rho^{(N)}(0)$.
Thereafter, the frequency to be estimated is encoded into the state of the probes; however, during the encoding procedure, which lasts
for a time $t$, the probes experience the action of some external noise so that they have to be treated as an open quantum system \cite{Breuer2002,Rivas2012}.
The state after the encoding is fixed by a completely positive
trace preserving linear map $\Lambda^{(N)}(t)$, via
\begin{equation}
\rho^{(N)}(t) = \Lambda^{(N)}(t)[\rho^{(N)}(0)].
\end{equation}
We focus on independent and identical noise, $\Lambda^{(N)}(t) =\Lambda(t) \otimes \ldots \otimes\Lambda(t)$, which yields a good approximation
in many circumstances and
is commonly considered in parameter estimation \cite{Demkowicz2015}.
After the encoding procedure, a measurement on the $N$ probes is performed to extract as much information about $\omega$ as possible.
The whole preparation-encoding-measurement protocol is repeated $\nu = T/t$ times, where $T$ is the total available time and we neglect
the preparation and measurement duration.

Based on the experimental data,
one defines an estimator, which is the random variable
giving the estimated value of $\omega$. By virtue of the (quantum) Cramer-Rao bound (CRB) \cite{Braunstein1994},
the estimation error, as quantified by the mean squared error of the estimator, is lower bounded by
\begin{equation}
\Delta^2 \omega(N)  \geq \min_{t \in [0,T]} \left(T F^{\omega}_Q[\rho^{(N)}(t)]/t\right)^{-1},
\end{equation}
for any initial state, evolution, measurement procedure and estimator, if the latter is consistent and unbiased.
The quantity $F^{\omega}_Q[\rho^{(N)}(t)]$ is the quantum Fisher information (QFI) of the probes' state with respect to $\omega$  after the encoding
and, indeed, it fixes the ultimate achievable precision; the CRB can be saturated in the limit of
an infinite number of repetitions, $\nu \rightarrow \infty$.
For high dimensional systems, it is in general very hard to compute the QFI, even numerically, but powerful analytical techniques \cite{Escher2011,Demkowicz2012} allow us to get some tight upper bound.

\subsection{The basic model}
In the basic model we exploit here
each probe is described as a two-level system and the noise acting on it is modelled
via a damped harmonic mode coupled to a zero temperature reservoir.
Accordingly, the evolution of the qubit-mode state $\rho_{qm}(t)$ is given by the Lindblad equation \cite{Breuer2002} ($\hbar = 1$)
\begin{eqnarray}
\frac{\mathd}{\mathd t} \rho_{qm}(t) &=& - i \left[\frac{\omega}{2} \sigma^z + \omega_m a^{\dag}a + \lambda H_I, \rho_{qm}(t)\right] \nonumber\\
&&+ \Gamma\left({a} \rho_{qm}(t) {a}^{\dag}-\frac{1}{2}\left\{{a}^{\dag} {a}, \rho_{qm}(t) \right\} \right), \label{eq:l1}
\end{eqnarray}
with $H_I = \sigma^- \otimes {a}^\dag+\sigma^+ \otimes {a}$. Here, $\omega$ is the frequency to be determined, $\sigma^+=\ketbra{1}{0}$ ($\sigma^-=\ketbra{0}{1}$)
is the raising (lowering) operator of the qubit and $\sigma^z=\ketbra{1}{1}-\ketbra{0}{0}$;
moreover, $\omega_m$, ${a}$ and ${a}^{\dag}$ are the frequency,
the annihilation operator and the creation operator of the mode, $\lambda$ is the qubit-oscillator coupling strength and $\Gamma$ is the oscillator's damping rate.
This system can also be understood as an instance of a spin-boson model for finite $T$
\cite{Lemmer2018}.
Finally, we assume that the initial qubit-mode state
is a product state with the mode in the vacuum state.
The dissipative dynamics
of the qubit state $\rho(t)=\mbox{Tr}_{m }\left\{\rho_{qm}(t)\right\} = \Lambda(t)[\rho(0)]$ induced by Eq.\eqref{eq:l1} is
then an amplitude damping,
where the excited-state population evolves as $\rho_{11} \mapsto \rho_{11}(t)=|f(t)|^2 \rho_{11}$,
while the coherence evolves as $\rho_{10} \mapsto \rho_{10}(t)=f(t) \rho_{10}$,
where
\begin{equation}\label{eq:fetiii}
f(t) = e^{- \left(i\omega+\frac{\chi}{4}\right) t}\left(\cosh\left(\Omega t /2 \right)+ \frac{\chi}{2 \Omega} \sinh\left(\Omega t /2\right)\right)
\end{equation}
with $\Delta = \omega-\omega_m$, $\chi = \Gamma- 2 i \Delta$ and $\Omega = \sqrt{\chi^2/4-4 \lambda^2}$ \cite{garraway1997,li2010}.
Since $|f(t)| \rightarrow 0$ for $t\rightarrow \infty$, there is no CT: the
interaction progressively destroys all the probes' coherences.

Upper bounding the QFI via the technique developed in \cite{Demkowicz2012}, along with the CRB, we can
lower bound the estimation error under amplitude damping with \cite{kolodynski2013}
\begin{equation}\label{eq:qcrbd2}
\Delta^2 \omega (N) \,  \geq \min_{t\in [0,T]}  \frac{1+\frac{N}{4} \left(|f(t)|^{-2}-1\right)}{N^2 T t}.
\end{equation}
From the previous relation, one can directly see that the SQL can be overcome asymptotically only if
we perform measurements on the short-time scale
(or in the presence of a full-revival in the dynamics: only in these cases
$f(t) \rightarrow 1$, so that the second term in the numerator will not dominate, and the SQL
will not be enforced, for $N\rightarrow\infty$
\cite{Smirne2016}).
Hence, expanding $f(t)$ in time and exploiting its quadratic decay for short times, one sees
that the optimal estimation time, minimizing the right hand side (r.h.s.) of Eq.\eqref{eq:qcrbd2}, for $N\rightarrow \infty$ is
$t_\mathrm{opt} = 2/(\lambda \sqrt{N}) + O\left(1/N \right)$,
from which
\begin{equation}\label{eq:qcrbt}
\Delta^2 \omega_{\text{ent}}(N) \,  \geq \frac{\lambda}{T N^{3/2}}.
\end{equation}
Indeed, the estimation strategy saturating this limit, i.e., yielding the asymptotically
optimal $N^{-3/2}$ scaling, requires initially entangled probes \cite{chin2012,Macieszczak2015,Smirne2016}.
We have thus shown for the model at hand how
the limits in the estimation precision due to the presence of noise \cite{huelga1997,Escher2011,Demkowicz2012}
can be at least partially avoided \cite{Smirne2016}.
Besides preparing entangled probes, one needs
to access the (Zeno \cite{Misra1977,Facchi2008}) short-time region of the dynamics,
where the temporal correlations of the noise induce deviations from a Lindblad evolution of the probes \cite{Breuer2002},
so that
the survival probabilities decay slower than exponentially;
even more, since $t_\mathrm{opt} \rightarrow 0$ for $N\rightarrow \infty$ (as $N^{-1/2}$),
measurements at a shorter and shorter times have to be performed when increasing $N$.

\begin{figure}[ht!]
\begin{center}
\includegraphics[width=0.7\columnwidth]{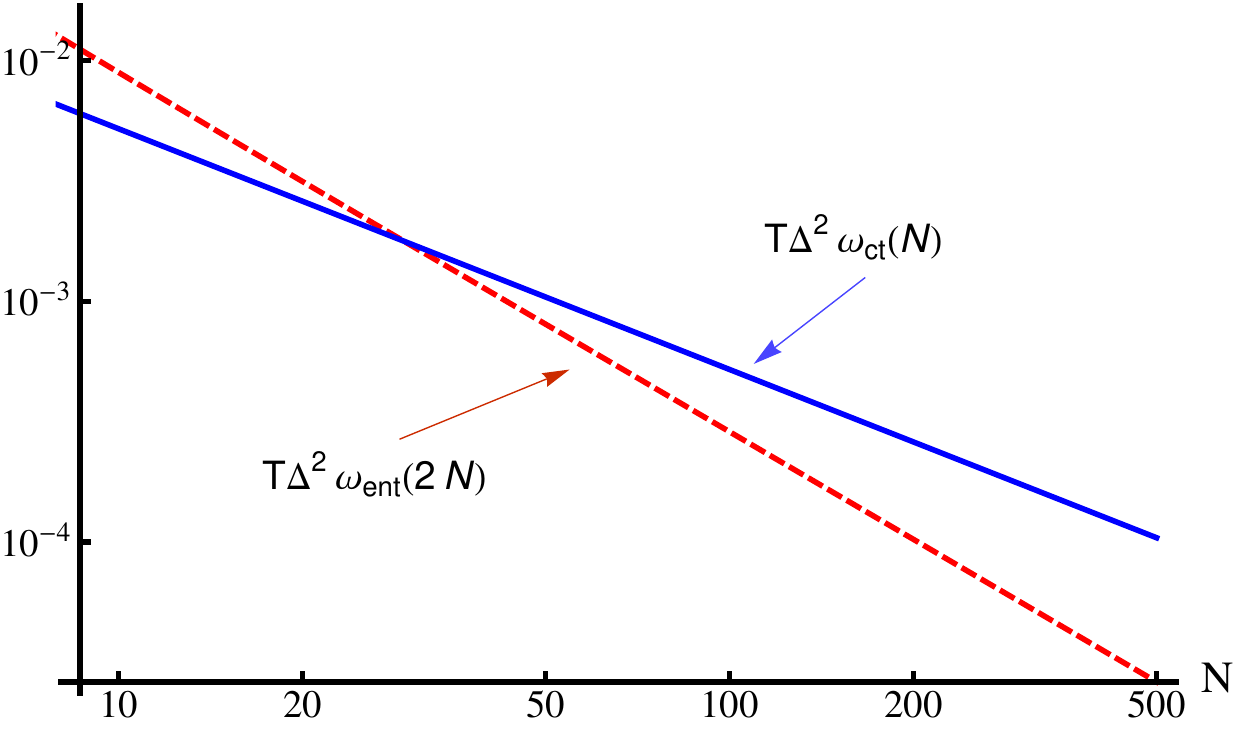}
\caption{Lower bound to any estimation error for the entanglement-based strategy obtained by minimizing the r.h.s. of Eq.\eqref{eq:qcrbd2} for $2N$ probes (red, dashed line)
and the estimation error in the Ramsey scheme for the CT-based strategy with $N$ probes (blue, solid line), see Eq.\eqref{eq:dctk}.
The parameters are $\bar{t}=30 \Gamma^{-1}, \Delta=0.05 \Gamma, \lambda=\tilde{\lambda}/2=0.3 \Gamma$.}
\label{fig:1}
\end{center}
\end{figure}

\section{Coherence trapping and enhancement of the estimation precision}%
Now, let us see how by modifying the
probes' dynamics we can devise a strategy, which not only allows to interrogate the probes in the
long-time regime, but can also lead to an enhancement of the estimation precision
for finite values of $N$.

For any qubit used as a probe, we couple a further ancillary qubit to the damped harmonic
oscillator.
The auxiliary qubit can be understood as a subset of the environmental degrees of freedom,
which are being manipulated to change the evolution of the probes
(at variance
with the ancilla-based schemes for metrology as those, e.g., in \cite{Demkowicz2014c},
where the dynamics is fixed and the ancillas do not interact with the environment).
Denoting as $\varrho(t)$ the two-qubit plus one mode state, we describe the global evolution with the Lindblad equation:
\begin{eqnarray}
\frac{\mathd}{\mathd t} \varrho(t) &=& - i \left[\frac{\omega}{2} \sigma^z+\frac{\tilde{\omega}}{2} \tilde{\sigma}^z + \omega_m a^{\dag}a + \lambda H_I + \tilde{\lambda} \tilde{H}_I, \varrho(t)\right] \nonumber\\
&&+ \Gamma\left({a} \varrho(t) {a}^{\dag}-\frac{1}{2}\left\{{a}^{\dag} {a}, \varrho(t) \right\} \right),\label{eq:l2}
\end{eqnarray}
where the tilde is used for parameters and operators referred to the ancilla qubit,
so that $\tilde{\sigma}^z \equiv \mathbbm{1}\otimes\sigma^z$ ($\sigma^z$ is used for $\sigma^z \otimes \mathbbm{1}$),
while $\tilde{H}_I = \tilde{\sigma}^- \otimes {a}^\dag+\tilde{\sigma}^+ \otimes {a}$; moreover,
we consider an initial product state, with at most one excitation overall.
The validity of this description depends on several conditions
involving the probe and ancilla interaction with the surrounding environment; in the next paragraph,
we will see how Eq.\eqref{eq:l2} does provide a satisfactory characterization for a relevant scenario
in ion traps.

The solution of the model can be directly obtained along the same line as \cite{garraway1997,lazarou2012,Man2012} and
is detailed in \cite{sup}.
The reduced dynamics of the probe qubit is still given by an amplitude damping, where the dissipative function
$f(t)$ has to be replaced by a different function $\tilde{f}(t)$, see \cite{sup} for its expression.
Under the resonant condition $\tilde{\omega} = \omega$ we do have the occurrence of CT:
in this case,
\begin{eqnarray}
\tilde{f}(t) = e^{-i \omega t}C_{\infty} \left[1+\frac{\lambda^2}{\tilde{\lambda}^2}  e^{-\frac{\chi t}{4}}
\left(\cosh\left(\frac{Z t}{2}\right)+\frac{\chi}{2Z} \sinh\left(\frac{Z t}{2}\right)\right)\right], \label{eq:ftil}
\end{eqnarray}
where
$
Z = \sqrt{\Omega^2-4 \tilde{\lambda}^2}
$
and
\begin{equation}
C_{\infty} = \frac{\tilde{\lambda}^2}{\lambda^2+\tilde{\lambda}^2}.
\end{equation}
The probes' coherences will thus (partially) survive the dynamics: $|\rho_{10}(t)|\rightarrow C_{\infty} |\rho_{10}|$
for $t\rightarrow \infty$, where $C_{\infty}>0$ for non-zero $\tilde{\lambda}$.

The basic idea to exploit CT for metrological purposes is now very simple.
If we can access the long-time region, where the coherence has practically stabilized to its asymptotic value,
the best estimation strategy will be to wait as long as possible:
due to CT, the effect of decoherence
has been essentially turned off, while the phase carrying information about $\omega$
keeps accumulating in the probes' state. 
For now, let us assume that we measure at a time $\bar{t}$, where
the coherence can be approximated with its asymptotic value.
In addition, we still assume 
$T/\bar{t} \gg 1$, so that we have a high number of
independent repetitions of the experiment,
justifying the statistical analysis we make here.
For the sake of concreteness, consider the Ramsey scheme.
One prepares the probes in an initial product state of identical
balanced superpositions of the ground and excited state, via a first $\pi/2$ pulse. After the evolution, which lasts for a time $\bar{t}$,
a second $\pi/2$ pulse is applied and then the excited state population is measured.
The resulting signal is given by
$\bar{P} \equiv P(\bar{t}) = \left(1+\mbox{Re}[\tilde{f}(\bar{t})]\right)/2$,
which, using $\tilde{f}(\bar{t}) \approx e^{-i \omega \bar{t}}C_{\infty}$,
can be approximated as $\bar{P}\approx \left(1+ C_{\infty} \cos(\omega \bar{t})\right)/2$.
The frequency uncertainty is \cite{huelga1997}
\begin{equation}
\Delta^2 \omega_{\text{ct}} (N)  = \frac{\bar{P}(1-\bar{P}) \bar{t}}{N T (\mathd \bar{P}/\mathd \omega)^2},
\label{eq:frequency_uncertainty}
\end{equation}
which is minimized for
$\omega \bar{t} = r \pi/2$,
for any odd $r$, so that the minimum error for the CT-based estimation strategy is
\begin{equation}\label{eq:dctk}
\Delta^2 \omega_{\text{ct}}(N) = \frac{C^{-2}_\infty}{N T \bar{t}}.
\end{equation}

Remarkably, the uncertainty in Eq.\eqref{eq:dctk} can be smaller than that in Eq.(\ref{eq:qcrbt}).
To get a more quantitative idea about the improvement due to CT with respect to the entangled-probe strategy,
consider the ratio
\begin{equation}\label{eq:nun}
G(N)=\frac{\Delta^2 \omega_{\text{ent}}(2N)}{\Delta^2 \omega_{\text{ct}}(N)}
= \frac{\lambda \bar{t} C^2_{\infty}}{\sqrt{8N}} = \frac{\lambda  \tilde{\lambda}^4 \bar{t}}{\sqrt{8N}(\lambda^2+\tilde{\lambda}^2)^2 }.
\end{equation}
Crucially, to
ensure a fair comparison between the two strategies, we took into account that CT requires one auxiliary qubit for any probe qubit, so
that for $N$ probes in the CT scheme, we use $2 N$ qubits in the entanglement-based scheme.
As expected, the gain due to CT will be the higher the bigger the asymptotic coherence, as well as the longer $\bar{t}$, while the entangled-probe strategy will always
be more precise for sufficiently large values of $N$.
For finite (and small) values of $N$, the lowest error in the entanglement-based strategy
may be smaller than that at the r.h.s. of Eq.\eqref{eq:qcrbt}; because of that, we also considered
the lower bound to the error which is given directly by the r.h.s. of Eq.\eqref{eq:qcrbd2}. 
The latter bounds the smallest estimation error with the entangled-probe strategy,
for any value of $N$ and measurement procedure (but without modifying the probes' evolution), since it is derived using the QFI.
The results are reported in Fig.\ref{fig:1}, where one can still clearly observe the transition between the regimes
where CT and entangled probes are the more accurate estimation strategy.

As a side remark, CT can occur for both
monotonic and oscillating decays of $|\tilde{f}(t)|$ to its asymptotic value.
For the amplitude damping, the (non-)monotonicity of $|\tilde{f}(t)|$ is equivalent to the (non-)Markovianity of the dynamics \cite{li2010}
(see \cite{Rivas2014,Breuer2016} for recent reviews on quantum Markovianity):
we conclude that non-Markovianity is not necessary to trigger the enhancement in the estimation precision pointed out in this paper.

\section{Ion-trap realization}%
In the following, we show how the CT estimation strategy for
enhanced precision in frequency estimation can be realized in an ion-trap setup.


Let us consider a trapped-ion qubit with an optical transition as a probe.
Since we need to implement spin-motion coupling on the probe transition, it is advantageous to work with an optical transition.
Let us further consider an ancilla ion
of the same species such that $\omega =\tilde{\omega}$ and that the ions form a Coulomb crystal.
One of the motional modes of the crystal can then be used to realize the dissipative mode. Damping on the mode can be implemented by laser cooling.
Since laser cooling is an incoherent process, it would compromise the desired internal state evolution if implemented via the probe
or ancilla ions. Hence, we need at least three ions to implement the scheme where the third ion provides cooling.

Ideally, the cooling ion has a mass very similar to the probe and ancilla species to provide effective cooling. At the same time,
the cooling transitions should be separated energetically as far as possible from the probe transition in order to avoid scattering
of photons from the cooling lasers by the probes and minimizing their ac-Stark shift on the probe transition.

For concreteness, we consider $^{40} \mathrm{Ca}^+$ as the probe and $^{24} \mathrm{Mg}^+$ as the coolant ion. There is a possible ``clock'' transition
near $729\,$nm in $^{40} \mathrm{Ca}^+$ between two states $\ket{0}$ and $\ket{1}$ from the $^2 S_{1/2}$ and $^2 D_{5/2}$ manifolds~\cite{schmidt_optical_clocks_review}.
$^{24} \mathrm{Mg}^+$, on the other hand, has been used for sympathetic cooling of mixed-species crystals before, with cooling transitions near $280\,$nm~\cite{nist_eit_cooling}.
Due to the large difference between the two transition frequencies, we neglect the influence of the cooling lasers on the qubit levels.

In the following, we assume that the ions form a Coulomb crystal along the trap axis of a linear Paul trap with harmonic confining potential. We consider
an arrangement $^{40} \mathrm{Ca}^+ - ^{40} \mathrm{Ca}^+ - ^{24} \mathrm{Mg}^+$ of the ions and focus on the axial motion. We assume that the ions are
sufficiently cold that their motion is described in terms of normal modes. The axial normal modes are well separated in frequency and we can use one
of the normal modes to realize the dissipative mode. We choose the highest frequency mode (the Egyptian mode in the case of a homogeneous crystal) as the dissipative mode.
For an axial trap potential where a single $^{40}\mathrm{Ca}^+$ ion has a center-of-mass frequency of $\omega_z/2\pi =1\,$MHz, the dissipative mode
has a frequency of $\omega_{3}/2\pi = 2.59\,$MHz.

Let us now analyze the time evolution of the ions in the trap. We assume that the motional degrees of freedom are cooled close the ground state initially
and that the $^{40}\mathrm{Ca}^+$ ions are initialized in the $\ket{0}$ state before the Ramsey sequence starts. Then, the probe ion is excited
to the state $|\psi_1 \rangle =\frac{1}{\sqrt{2}}(\ket{0} +\mathrm{i} \ket{1})$ by the first Ramsey pulse from the probe laser. Note that we assume that the
state of the ancilla qubit remains $\ket{0}$. During the subsequent free evolution time, the system of probe and ancilla ions and dissipative mode should
evolve according to Eq.~\eqref{eq:l2}. The coherent part of this time evolution can be realized by illuminating the ions with a laser tuned to the
first red-sideband transition of the $^{40}\mathrm{Ca}^+$ ions and the dissipative mode, see~\cite{sup}. The dissipative mode is sympathetically cooled through
the $^{24}\mathrm{Mg}^+$ ion. Here, we assume that EIT cooling~\cite{morigi_laser_cooling} is realized as it allows for high cooling rates at
relatively small laser powers~\cite{nist_eit_cooling}.
However, laser cooling is not described by the dissipator in Eq.~\eqref{eq:l2} because it does not perfectly realize coupling to a zero temperature reservoir.
The dissipator for laser cooling reads~\cite{cirac_laser_cooling, morigi_laser_cooling}
\begin{equation}
\fl \mathcal{D}_\mathrm{lc} \varrho (t) = \Gamma(\bar{n}+1) \left[ a \varrho(t) a^{\dagger}-\frac{1}{2}\{ a^{\dagger} a, \varrho(t)\} \right]
+ \Gamma \bar{n} \left[ a^{\dagger} \varrho(t) a- \frac{1}{2} \{a a^{\dagger}, \varrho(t) \} \right],
\label{eq:laser_cooling_dissipator}
\end{equation}
where $\Gamma$ is the cooling rate and ``lc'' stands for ``laser cooling''. $\bar{n}$ is the final occupation number of the thermal state of
the dissipative mode if subject to the above dissipator. For a realistic assessment of the protocol, we also have to include the finite linewidth
of the transition in the probe and ancilla ions. To this end, we have to complement $\mathcal{D}_\mathrm{lc}$ with
\begin{equation}
\fl \mathcal{D}_\mathrm{se} \varrho (t) = \Gamma_\mathrm{se} \left[\sigma^- \varrho (t) \sigma^+ -\frac{1}{2}\{\sigma^+ \sigma^-, \varrho(t) \} \right] + \Gamma_\mathrm{se} \left[\tilde{\sigma}^- \varrho (t) \tilde{\sigma}^+ -\frac{1}{2}\{\tilde{\sigma}^+ \tilde{\sigma}^-, \varrho(t) \} \right].
\label{eq:def_d_se}
\end{equation}
Hence, in a trapped ion experiment the system evolves according to Eq.~\eqref{eq:l2} with the dissipator replaced by
$\mathcal{D}_\mathrm{lc} + \mathcal{D}_\mathrm{se}$ during the free evolution period.
Finally, the experimental cycle is completed by a second $\pi/2$-pulse and a measurement of $\sigma^z$ on the probe qubit.

Let us now proceed to show that we can indeed obtain an advantage in precision over the best entangled strategy, i.e. $G(N)>1$,
in an ion trap experiment. To this end, we simulate trapped-ion experiments with realistic parameters in order to show that the advantage persists
also in this case for appropriate parameters. The crucial difference between the trapped-ion realization and the scenario considered in the first part
is that we cannot attain zero temperature for the dissipative mode in a real experiment.

In the simulations, we assume an initial product state of qubits and mode with the probe and ancilla qubits in state $\ket{0}$ and the mode in a thermal state
with variable mean occupation number $\bar{n}$. We truncate the motional Hilbert space at $n_\mathrm{max} = 7$ excitations. The spontaneous emission rate
of the considered transition in $^{40}\mathrm{Ca}^+$ is $\Gamma_\mathrm{se}/2\pi=0.14\,$Hz~\cite{schmidt_optical_clocks_review}.
In this setting we compute the evolution of the full system up to a final time $\Gamma \bar{t} = 180$ for $N_{\omega} = 100$ equally spaced values
of $\omega \in [-0.1,0.1]\,\mathrm{kHz}$ with $\Gamma/2\pi=1\,\mathrm{kHz},\:\lambda/2\pi=0.1\,\mathrm{kHz},\:\tilde{\lambda}/2\pi=-0.29\,\mathrm{kHz}$
and $\omega_\mathrm{m} = 0$ fixed. Note that we can set $\omega_\mathrm{m}=0$ without loss of generality, see~\cite{sup}.
Finally, we compute the uncertainty in the estimated frequency according to Eq.~\eqref{eq:frequency_uncertainty}.
Note that we take the first $^{40}\mathrm{Ca}^+$ ion to be the probe ion and the middle $^{40}\mathrm{Ca}^+$ ion
as the ancilla. The ratio of the normalized amplitudes of the dissipative mode at these positions is about $-2.9$ such that we assume the ions are
illuminated with equal intensity.

\begin{figure}
\includegraphics[width=0.48\columnwidth]{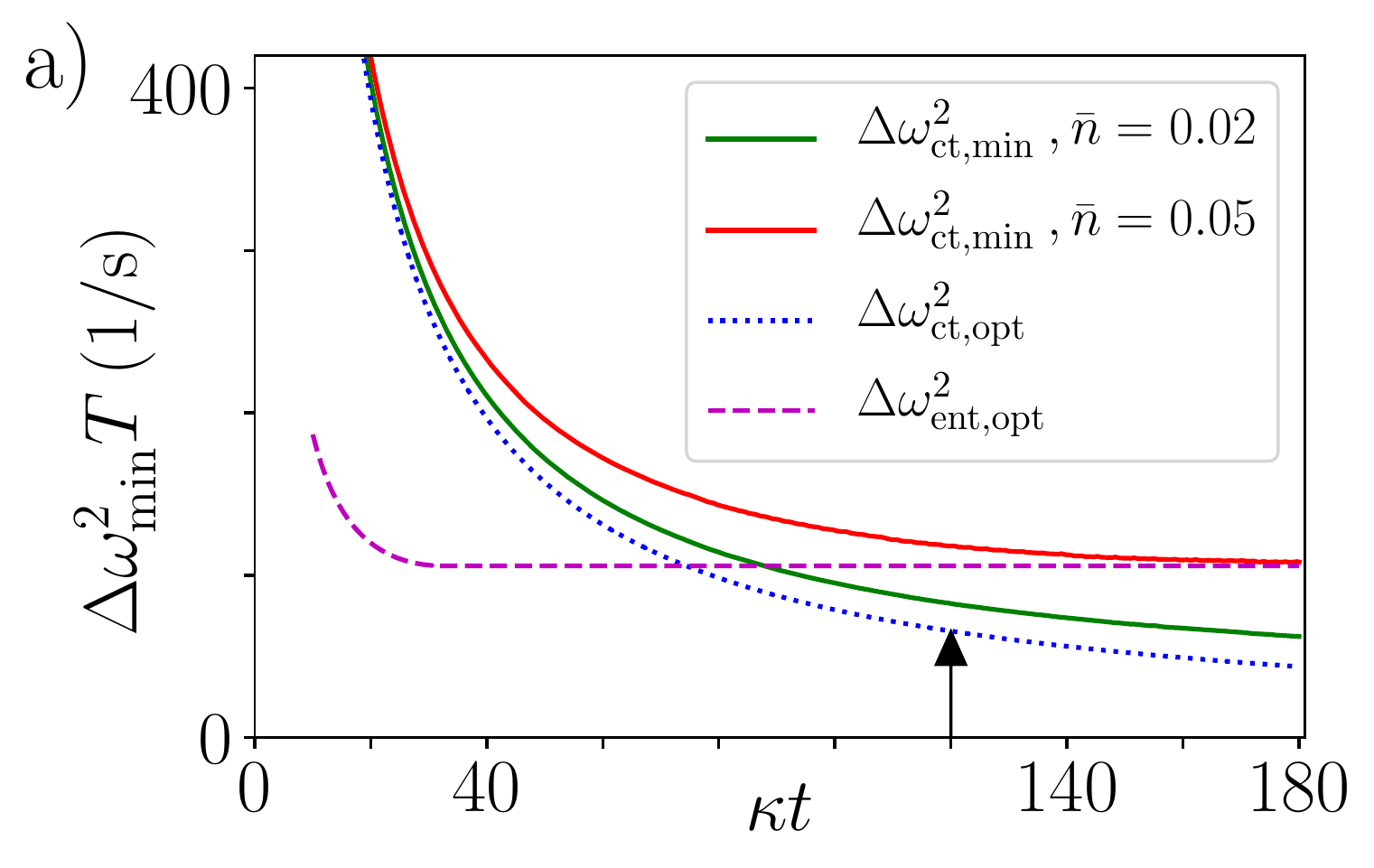}
\includegraphics[width=0.48\columnwidth]{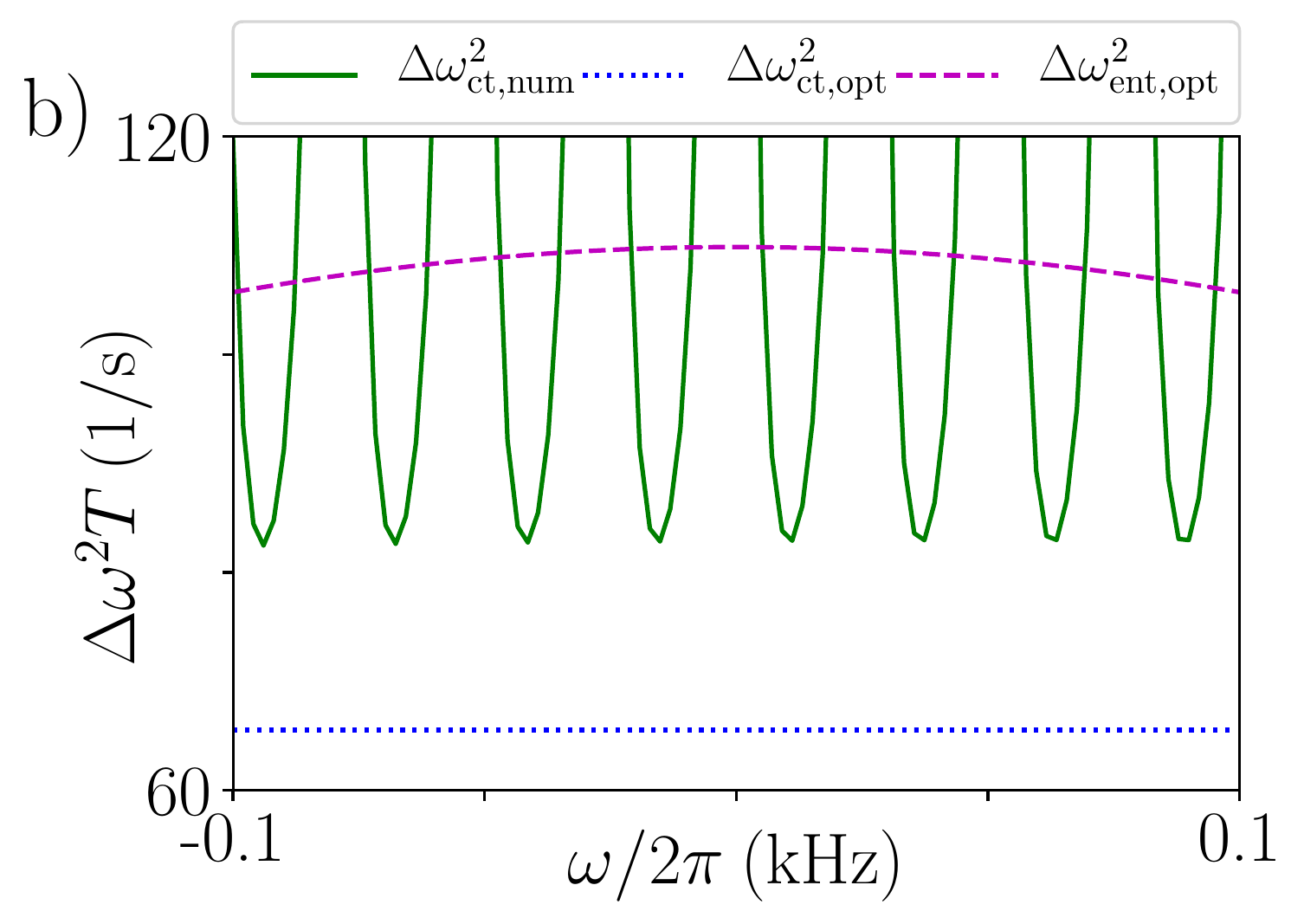}
\caption{Frequency uncertainty of the CT parameter estimation strategy. \textbf{a)} The figure shows the minimal uncertainty $\Delta \omega^2 T$ for the
CT strategy for environments with different temperatures. In the case of zero temperature, we obtain the uncertainty of~\eqref{eq:dctk}. The figure
also shows the minimal uncertainty for the entangled strategy. For times $t > \kappa/100$ the CT strategy outperforms the entangled strategy. Part \textbf{b)}
shows the uncertainty as a function of $\omega$ for $\bar{t} \approx 19.1\,$ms. The point is indicated by the arrow in part \textbf{a)}. }
\label{fig:experimental_sensitivity}
\end{figure}

In Fig.~\ref{fig:experimental_sensitivity} we show the results of our simulations. Part \textbf{a)} of the figure depicts the minimal uncertainty $\Delta \omega^2$ of the CT strategy
multiplied by the total time $T$ as a function of time. For every value of $t$, we show the minimal uncertainty for $\omega \in [-0.1,0.1]\,\mathrm{kHz}$.
The plot depicts the results of the CT strategy for finite temperature reservoirs with $\bar{n}=0.02$ and $\bar{n}=0.05$ as well as the zero temperature limit, Eq.~\eqref{eq:dctk}.
For comparison, the figure also shows the minimum of Eq.~\eqref{eq:qcrbd2}, i.e. the minimum uncertainty for the entangled strategy, for the considered parameters.
For times $\bar{t} > 100/\kappa \approx 15.9\,$ms the CT strategy outperforms the entangled strategy.
Clock laser coherence times of $\approx 300\,$ms have been reported~\cite{huntemann_hyperramsey} so that one should be able to reach this time scale in practice.
Note that for the entanglement-based
strategy we have considered the best possible achievable precision,
using the lower bound in Eq.\eqref{eq:qcrbd2} not including
any additional experimental noise (e.g., due to thermal effects). The figure illustrates that an increasing temperature reduces the gain in precision by the CT
strategy for fixed $\lambda$ and $\tilde{\lambda}$. We note that increasing the ratio $\tilde{\lambda}/\lambda$ it is also possible to obtain a gain in precision for higher
temperatures. However, for this we would need a higher laser intensity at the ancilla ion than at the probe ion.

In part \textbf{b)} of the figure we show the uncertainty as a function of the probe frequency for a fixed $\bar{t} = 120/\kappa \approx 19.1\,$ms.
Accordingly, the condition $\omega \bar{t} = r \pi/2$, where $r$ is odd, cannot be satisfied for all $\omega$ and, in particular, for some $\omega$ division by small numbers occurs
in Eq.~\eqref{eq:frequency_uncertainty}.

\section{Conclusions}
We have described a strategy to estimate the value of an unknown frequency
based on CT, that is, on the presence of quantum coherence in the state of the sensing
particles on long time scales, despite the presence of noise.
CT frequency estimation does not require entanglement between the probes
or measurements on short time scales while it can outperform the best
entanglement-based strategy when dealing with a small number of probes.
This was shown by taking into account a qubit interacting with
a damped harmonic oscillator, as well as an ion trap configuration
under achievable conditions for current technology.
Our approach thus paves the way for a deeper investigation of parameter estimation
relying on a realistic description and full exploitation of the finite resources at disposal.

\ack
We would like to thank Rafa\l{} Demkowicz-Dobrza\'{n}ski, Jan Haase and Jan Ko\l{}ody\'{n}ski
for many useful discussions.
We acknowledge financial support by the ERC Synergy Grant BioQ (grant no 319130)
and the EU project QUCHIP (grant no. 641039)
\\
\textit{Note Added:} During the completion of this paper, the related works \cite{Guo2017,Wang2017}
appeared, in which an asymptotic non-zero value of the QFI has been shown to occur for different
open-system dynamics.

\appendix
\section{Explicit solution of the model with an auxiliary qubit}
Given the master equation \eqref{eq:l2},
if the total initial state $\varrho(0)$ has at most one excitation,
the resulting dynamics is formally equivalent to the evolution of a two-level system interacting with a damped mode, in turn coupled
to an undamped mode (both the couplings being in the Jaynes-Cummings form and with at most one initial excitation).
The latter model was shown to describe properly the dynamics induced by a band-gap spectral density \cite{garraway1997}
and it is known that the resulting dynamics of the (two-level) probe state, $\rho(t)=\mbox{Tr}_{m \tilde{q}}\left\{\varrho(t)\right\}$ (where $\mbox{Tr}_{m \tilde{q}}\left\{\cdot\right\}$
denotes the partial trace over the mode and the auxiliary qubit),
exhibits both population \cite{garraway1997,lazarou2012} and coherence \cite{man2015} trapping, under specific resonance conditions.
We report here the solution of the model, for the sake of self-consistency.

If we start from a state with at most one excitation, given that the considered Hamiltonians preserve the excitation number
and we neglect any absorption process,
we can make the ansatz that the state at time $t$ is of the form \cite{garraway1997}
\begin{equation}\label{eq:anz}
\varrho(t) = \Pi(t) \ket{000}\bra{000} + \ket{\psi (t)}\bra{\psi(t)},
\end{equation}
for a certain function $\Pi(t)$ and where $\ket{\psi(t)}$ is the non-normalized vector given by
\begin{equation}\label{eq:anz2}
 \ket{\psi (t)} = \kappa(t) \ket{100}+a(t)\ket{010}+\tilde{\kappa}(t)\ket{001}+w \ket{000},
\end{equation}
where now $\ket{100}, \ket{010}$ and $\ket{001}$ denote the pure state with one excitation in, respectively, the probe-qubit, the mode and
the ancilla-qubit; indeed $\ket{000}$ is the vacuum state.
Replacing Eqs.(\ref{eq:anz}) and (\ref{eq:anz2}) in the master equation \eqref{eq:l2},
one obtains that the dynamics is equivalently described by the following system of equations for the probability amplitudes of
qubit, ancilla and mode excitation, respectively:
\begin{eqnarray}
i \dot{\kappa}(t) &=& \omega \kappa(t) + \lambda a(t) \nonumber\\
i \dot{\tilde{\kappa}}(t) &=& \tilde{\omega} \tilde{\kappa}(t) + \tilde{\lambda} a(t) \nonumber\\
i \dot{a}(t) &=& \left(\omega_m-i \frac{\Gamma}{2}\right) a(t)+\lambda \kappa(t)+\tilde{\lambda}\tilde{\kappa}(t).
\end{eqnarray}
This can be formally solved moving to the frequency domain: for the initial conditions
\begin{eqnarray}
\kappa(0) &=& \kappa_0 \nonumber\\
\tilde{\kappa}(0) &=& a(0) = 0
\end{eqnarray}
and denoting as $\mathcal{I}[g(z)](t)$ the inverse Laplace transform of the function $g(z)$ evaluated in $t$, one finds
\begin{equation}
\fl \frac{\kappa(t)}{\kappa_0} = \mathcal{I}\left[\frac{\tilde{\lambda}^2+(z+ i \bar{z})(z+i \tilde{\omega})}{z^3+i z^2 (\bar{z}+\omega+\tilde{\omega})
+z(\tilde{\lambda}^2+\lambda^2-\bar{z}(\omega+\tilde{\omega})-\omega \tilde{\omega})
+i(\tilde{\lambda}^2 \omega+\lambda^2 \tilde{\omega}-\bar{z}\omega \tilde{\omega})}\right](t), \label{eq:ilt}
\end{equation}
where
\begin{equation}
\bar{z} = \omega_m-i \frac{\Gamma}{2}.
\end{equation}

Given the global state in Eq.(\ref{eq:anz}), one can easily see that the corresponding reduced dynamics
of the probe-qubit is an amplitude damping dynamics,
fixed by the function $\tilde{f}(t) = \kappa(t)/\kappa_0$.
In particular, for $\omega=\tilde{\omega}$, the inverse Laplace transform in Eq.\eqref{eq:ilt} can be performed explicitly, getting Eq.\eqref{eq:ftil}.

\section{Derivation of Liouvillian of CT dynamics in an ion trap}

Here, we will show how we can obtain a system that evolves according to Eq.~\eqref{eq:l2} in an ion trap setup.
In the main text, we consider a three-ion Coulomb crystal along the trap axis of a linear Paul trap with harmonic confining potential.
We choose our coordinate system such that the $z$-axis coincides with the trap axis and consider an arrangement
$^{40} \mathrm{Ca}^+ - ^{40} \mathrm{Ca}^+ - ^{24} \mathrm{Mg}^+$ of the ions.
Furthermore, we consider that the ions are sufficiently cold that their motion can be described in terms of a set of $N_\mathrm{m}=3$ normal modes in each
direction~\cite{james_normal_modes, morigi_inhom_crystals}.

Let us now focus on the motion in the axial direction. The motional Hamiltonian in $z$ reads
\begin{equation}
H_\mathrm{m} = \sum_{n=1}^{N_\mathrm{m}} \omega_n a_n^{\dagger} a_n,
\end{equation}
where $a_n$ ($a_n^{\dagger}$) and $\omega_n$ are the annihilation (creation) operator and the frequency of mode $n$, respectively.
For an axial trap potential where a single $^{40}\mathrm{Ca}^+$ ion has a center-of-mass frequency of $\omega_z/2\pi =1\,$MHz, the axial normal mode frequencies
are $(\omega_1,\omega_2,\omega_3) = 2\pi(1.06,1.95,2.59)\,\mathrm{MHz}$.
Hence, the normal modes are indeed well-separated in frequency.
Recall that we choose the highest frequency mode $(n=3)$ as the dissipative mode.
The internal levels of the $^{40} \mathrm{Ca}^+$ ions are described by the Hamiltonian
\begin{equation}
H_{\mathrm{int}} =\sum_{j=1,2}\frac{\omega}{2}\sigma_j^z
\end{equation}
while those of $^{24}\mathrm{Mg}^+$ are adiabatically eliminated in the description of EIT cooling~\cite{morigi_laser_cooling}.

Let us now see how we can achieve the time evolution of Eq.~\eqref{eq:l2} during the free time evolutions. EIT cooling on the dissipative mode
through $^{24}\mathrm{Mg}^+$ leads to the dissipator in Eq.~\eqref{eq:laser_cooling_dissipator} of the main text~\cite{cirac_laser_cooling, morigi_laser_cooling}
\begin{equation}
\fl \mathcal{D}_\mathrm{lc} \varrho (t) = \Gamma(\bar{n}+1) \left[ a_3 \varrho(t) a_3^{\dagger}-\frac{1}{2}\{ a_3^{\dagger} a_3, \varrho(t)\} \right] + \Gamma \bar{n} \left[ a_3^{\dagger} \varrho(t) a_3- \frac{1}{2} \{a_3 a_3^{\dagger}, \varrho(t) \} \right],
\label{eq:laser_cooling_dissipator_app}
\end{equation}
where $\Gamma$ is the cooling rate and $\bar{n}$ is asymptotic occupation number of the thermal state that the dissipator takes the mode into.
The interaction of the $^{40} \mathrm{Ca}^+$ ions and the laser that is tuned to the first red-sideband transition of the dissipative mode is
described by the interaction Hamiltonian
\begin{equation}
H_{\mathrm{I}} =
\sum_{j=1,2} \frac{\Omega_j}{2} \mathrm{e}^{\mathrm{i} {\bf k}_\mathrm{L} \cdot {\bf r}_j} \mathrm{e}^{\mathrm{i} \phi_j} \sigma_j^+ \mathrm{e}^{-\mathrm{i} \omega_{\rm L} t} + \mathrm{H.c.},
\label{laser_ion_int_ham}
\end{equation}
where $\Omega_j (\phi_j)$ is the laser Rabi frequency (phase) at ion $j$ located at ${\bf r}_j$. $\omega_{\rm L}$ and ${\bf k}_\mathrm{L}$ are the laser frequency and
wave vector, respectively. Note that we have performed a rotating wave approximation using $\Omega_j \ll \omega_{\rm L}$ here.

The laser frequency can be written as $\omega_{\rm L} = \omega - \omega_3 + \delta$, where $\delta \ll \omega_3$. Assuming the Lamb-Dicke regime,
we can expand the exponentials $\mathrm{e}^{\mathrm{i} {\bf k} \cdot {\bf r}_j}$ in the Hamiltonian of Eq.~\eqref{laser_ion_int_ham} to first order in the Lamb-Dicke factors
$\eta_{jn} = k_z\sqrt{\hbar/(2 m_j \omega_n)} \ll 1$, where $k_z$ is the $z$-component of $\textbf{k}$.
For $\Omega_j \ll \omega_3$ and $\eta_{jn}\Omega_j \ll \omega_3-\omega_{1/2}$, we can neglect all terms in $H_{\mathrm{I}}$ except for the coupling of the spin to the dissipative mode.
In an interaction picture with respect to $H_0=H_\mathrm{m}+H_\mathrm{int}$, the interaction Hamiltonian can then be written as
\begin{equation}
H_{\rm I} = \sum_{j} (\lambda_j \sigma_j^+ a_3  \mathrm{e}^{-\mathrm{i} \delta t} + \mathrm{H.c.}),
\end{equation}
where $\lambda_j = \mathrm{i} \tilde{B}_{j3} \eta_{j3} \Omega_j \mathrm{e}^{\mathrm{i} (k_z z_j + \phi_j)} /2$. Here, $\tilde{B}_{j3}$ is the amplitude of the dissipative mode
mode at ion $j$ in mass-weighted coordinates~\cite{morigi_inhom_crystals}.

Now, moving to a second interaction picture with respect to the Hamiltonian $\tilde{H}_0 = -\sum_j (\tilde{\omega}_0/2)\sigma_j^z - \omega_{\rm m} a_3^{\dagger} a_3 $,
where $ \tilde{\omega}_0-\omega_{\rm m} = -\delta$, and setting $a_3\equiv a$ for clarity, we obtain
\begin{equation}
\tilde{H}_{\rm I} = \sum_j \frac{\tilde{\omega}_0}{2}\sigma_j^{z}+  \omega_{\rm m} a^{\dagger} a +  \sum_{j} (\lambda_j \sigma_j^+ a + \mathrm{H.c.})
\label{eq:effective_ion_ham}
\end{equation}
Writing $\sigma_1^z = \sigma^z,\:\sigma_2^z=\tilde{\sigma}^z,\dots $, this is exactly the Hamiltonian in Eq.~\eqref{eq:l2}. Taking into account the laser
cooling, the system evolves according to
\begin{equation}
\frac{\mathrm{d}}{\mathrm{d}t}\varrho(t) = -\mathrm{i}[\tilde{H}_{\rm I},\varrho(t) ] + \mathcal{D}_\mathrm{lc} \varrho(t).
\label{eq:full_time_ev_ions}
\end{equation}
For $\bar{n}=0$, Eq.~\eqref{eq:full_time_ev_ions} reduces to Eq.~\eqref{eq:l2}.
Note that the frequencies $\tilde{\omega}_0,\omega_{\rm m}$ in Eq.~\eqref{eq:effective_ion_ham} are in principle arbitrary in the chosen picture
and therefore we can set $\omega_{\rm m}= 0$ without loss of generality as we do in the main text. $\tilde{\omega}_0$ then corresponds to the detuning
of the laser from resonance assuming that the motional frequency is known.

Finally, if we include spontaneous emission from the probe transition, the dissipator in Eq.~\eqref{eq:full_time_ev_ions} becomes
$\mathcal{D}_\mathrm{lc} \to \mathcal{D} = \mathcal{D}_\mathrm{lc} + \mathcal{D}_\mathrm{se}$, where
\begin{equation}
\fl \mathcal{D}_\mathrm{se} \varrho (t) = \Gamma_\mathrm{se} \left[\sigma^- \varrho (t) \sigma^+ -\frac{1}{2}\{\sigma^+ \sigma^-, \varrho(t) \} \right] + \Gamma_\mathrm{se} \left[\tilde{\sigma}^- \varrho (t) \tilde{\sigma}^+ -\frac{1}{2}\{\tilde{\sigma}^+ \tilde{\sigma}^-, \varrho(t) \} \right].
\label{eq:def_d_se}
\end{equation}

\end{document}